\documentclass[twoside,twocolumn,9pt]{article}
\usepackage{extsizes}
\usepackage[super,sort&compress,comma]{natbib} 
\usepackage[version=3]{mhchem}
\usepackage[left=1.5cm, right=1.5cm, top=1.785cm, bottom=2.0cm]{geometry}
\usepackage{balance}
\usepackage{widetext}
\usepackage{times,mathptmx}
\usepackage{sectsty}
\usepackage{graphicx} 
\usepackage{lastpage}
\usepackage[format=plain,justification=raggedright,singlelinecheck=false,font={stretch=1.125,small,sf},labelfont=bf,labelsep=space]{caption}
\usepackage{float}
\usepackage{fancyhdr}
\usepackage{fnpos}
\usepackage[english]{babel}
\usepackage{array}
\usepackage{droidsans}
\usepackage{charter}
\usepackage[T1]{fontenc}
\usepackage[usenames,dvipsnames]{xcolor}
\usepackage{setspace}
\usepackage[compact]{titlesec}
\usepackage{epstopdf}%This line makes .eps figures into .pdf - please comment out if not required.

\definecolor{cream}{RGB}{222,217,201}

\begin{document}

\pagestyle{fancy}
\thispagestyle{plain}

%%%FIGURE SETUP - please do not change any commands within this section%%%
\makeatletter 
\newlength{\figrulesep} 
\setlength{\figrulesep}{0.5\textfloatsep} 

\newcommand{\topfigrule}{\vspace*{-1pt}% 
\noindent{\color{cream}\rule[-\figrulesep]{\columnwidth}{1.5pt}} }

\newcommand{\botfigrule}{\vspace*{-2pt}% 
\noindent{\color{cream}\rule[\figrulesep]{\columnwidth}{1.5pt}} }

\newcommand{\dblfigrule}{\vspace*{-1pt}% 
\noindent{\color{cream}\rule[-\figrulesep]{\textwidth}{1.5pt}} }

\makeatother
%%%END OF FIGURE SETUP%%%

%%%TITLE, AUTHORS AND ABSTRACT%%%
\twocolumn[
  \begin{@twocolumnfalse}
\vspace{3cm}
\sffamily
\begin{tabular}{m{4.5cm} p{13.5cm} }

 & \noindent\LARGE{\textbf{Anisotropic thermal conductivity and mechanical properties of phagraphene: A molecular dynamics study}} \\%Article title goes here instead of the text "This is the title"
\vspace{0.3cm} & \vspace{0.3cm} \\

 & \noindent\large{Luiz Felipe C. Pereira\textit{$^{a}$}, Bohayra Mortazavi\textit{$^{b}$}, Meysam Makaremi\textit{$^{c}$} and Timon Rabczuk\textit{$^{d}$} } \\%Author names go here instead of "Full name", etc.

\\

 & \noindent\normalsize{Phagraphene is a novel 2D carbon allotrope with interesting electronic properties which has been recently theoretically proposed. Phagraphene is similar to a defective graphene structure with an arrangement of pentagonal, heptagonal and hexagonal rings. In this study we investigate thermal conductivity and mechanical properties of phagraphene using molecular dynamics simulations. Using the non-equilibrium molecular dynamics method, we found the thermal conductivity of phagraphene to be anisotropic, with room temperature values of $218 \pm 20$ W/m-K along the armchair direction and $285 \pm 29$ W/m-K along the zigzag direction. Both values are one order of magnitude smaller than pristine graphene.
Analysis of phonon group velocities also shows a significant reduction in this quantity for phagraphene in comparison to graphene.
By performing uniaxial tensile simulations, we studied the deformation process and mechanical response of phagraphene. We found that phagraphene exhibits a remarkable high tensile strength around $85 \pm 2$ GPa, whereas its elastic modulus is also anisotropic along in-plane directions, with values of $870 \pm 15$ GPa and $800 \pm 14$ GPa for armchair and zigzag directions respectively.
The lower thermal conductivity of phagraphene along with its predicted electronic properties suggests that it could be a better candidate than graphene in future carbon-based thermoelectric devices.} 
\\%The abstract goes here instead of the text "The abstract should be..."

\end{tabular}

 \end{@twocolumnfalse} \vspace{0.6cm}

  ]
%%%END OF TITLE, AUTHORS AND ABSTRACT%%%

%%%FONT SETUP - please do not change any commands within this section
\renewcommand*\rmdefault{bch}\normalfont\upshape
\rmfamily
\section*{}
\vspace{-1cm}

%%%FOOTNOTES%%%

\footnotetext{\textit{$^{a}$~Departamento de F\'{\i}sica Te\'orica e Experimental, Universidade Federal do Rio Grande do Norte, 59078-970 Natal, Brazil; E-mail: pereira@fisica.ufrn.br}}
\footnotetext{\textit{$^{b}$~Institute of Structural Mechanics, Bauhaus-Universit\"at Weimar, Marienstr. 15, D-99423 Weimar, Germany; E-mail: bohayra.mortazavi@gmail.com}}
\footnotetext{\textit{$^{c}$~Chemical Engineering Department, Carnegie Mellon University, Pittsburgh, Pennsylvania 15213, USA.}}
\footnotetext{\textit{$^{d}$~Institute of Structural Mechanics, Bauhaus-Universit\"at Weimar, Marienstr. 15, D-99423 Weimar, Germany; E-mail: Timon.rabczuk@uni-weimar.de}}

%Please use \dag to cite the ESI in the main text of the article.
%If you article does not have ESI please remove the the \dag symbol from the title and the footnotetext below.
%\footnotetext{\dag~Electronic Supplementary Information (ESI) available: [details of any supplementary information available should be included here]. See DOI: 10.1039/b000000x/}
%additional addresses can be cited as above using the lower-case letters, c, d, e... If all authors are from the same address, no letter is required

%\footnotetext{\ddag~Additional footnotes to the title and authors can be included \emph{e.g.}\ `Present address:' or `These authors contributed equally to this work' as above using the symbols: \ddag, \textsection, and \P. Please place the appropriate symbol next to the author's name and include a \texttt{\textbackslash footnotetext} entry in the the correct place in the list.}

%%%END OF FOOTNOTES%%%

%%%MAIN TEXT%%%%

\section{Introduction}

Since the first experiments with graphene \cite{Novoselov2004, Novoselov2005, Geim2007} the range of predicted and produced two-dimensional carbon-based materials has increased considerably \cite{Li2014b}. 
Graphene remains the most stable planar form of carbon with honeycomb atomic arrangement. Nonetheless, some of these novel carbon-based materials have already been produced, such as nitrogenated holey graphene \cite{Mahmood2015} and amorphized graphene heterostructures \cite{Kotakoski2015}, while some have only been theoretically proposed, such as the carbon allotropes penta-graphene \cite{Zhang2015a} and phagraphene \cite{Wang2015a}.
Phagraphene is a new planar carbon allotrope composed of 5-6-7 carbon rings \cite{Wang2015a}. First principles calculations based on density functional theory have shown that this planar carbon allotrope is energetically comparable to graphene \cite{Wang2015a}. 
In fact, these recent first principles calculations confirm that phagraphene is more favorable than other planar carbon allotropes proposed so far, not only because of its sp$^2$-hybridization, but also due to its dense  atomic packing \cite{Wang2015a}. 

Graphene presents an extraordinary and unique combination of superior physical properties, such as ballistic electronic transport \cite{CastroNeto2009, Yazyev2010}, high thermal conductivity \cite{Balandin2008, Xu2014, Fugallo2014} and remarkable mechanical strength \cite{Lee2008}. 
Among the novel carbon-based materials there is a wealth of physical properties. For example, nitrogenated holey graphene is a semiconductor with a band gap $\approx 1.96$ eV  \cite{Mahmood2015}, while penta-graphene is predicted to have a $3.25$ eV band gap \cite{Zhang2015a}.
Similar to graphene, phagraphene also presents peculiar electronic properties due to the prediction of distorted Dirac cones. 
Nevertheless, to the best of our knowledge, the thermal conduction and mechanical properties of phagraphene have not yet been studied neither theoretically nor experimentally. 
The objective of this work is therefore to evaluate thermal transport properties and mechanical strength of phagraphene via computer simulations in order to guide future experimental investigations.

In the present work we employ classical molecular dynamics simulations to calculate the thermal conductivity of phagraphene at room temperature, as well as to determine its elastic modulus and mechanical strength.
The in-plane thermal conductivity of phagraphene was obtained via non-equilibrium molecular dynamics simulations (NEMD), which include a finite temperature gradient and resemble most experimental setups \cite{Xu2014}.
The mechanical response of phagraphene was investigated by performing uniaxial tensile stress simulations, from which we determine the elastic modulus and the breaking strength of the material.
We find that the thermal conductivity of phagraphene is one order of magnitude smaller than that of graphene, while its elastic modulus and tensile strain are comparable to graphene, albeit a little smaller.
Remarkably, our simulations show that the thermal conductivity and the elastic modulus of phagraphene are direction dependent, presenting an in-plane anisotropy not observed in the case of graphene and other two-dimensional carbon-based materials.

\section{Molecular dynamics modeling }

\begin{figure}[htb]
\centering
\includegraphics[width=\linewidth]{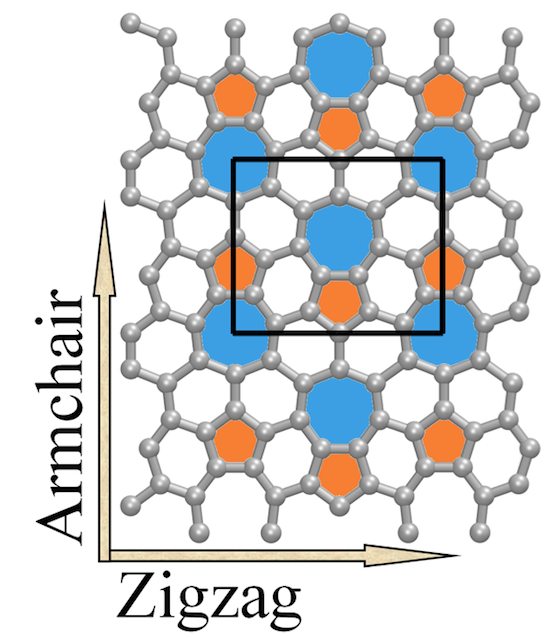}
\caption{A periodic supercell in the atomic structure of phagraphene. A 20-atom unit cell is also shown. In this study the properties are investigated along armchair and zigzag directions.}
\label{fig01}
\end{figure}

The atomic structure of phagraphene is depicted in Fig. \ref{fig01}, which includes a special pattern of pentagonal,  hexagonal and heptagonal carbon rings, as well as space-inversion symmetry. In fact, there are only 6 inequivalent carbon atoms in the 20-atom rectangular unit cell shown in Fig. \ref{fig01}.
Also indicated are the armchair and zigzag directions of the phagraphene lattice structure, defined in analogy to the ones in graphene.
In our study, both mechanical and heat conduction properties were investigated along armchair and zigzag directions. 
All structures simulated in this work are periodic along the planar directions. Interaction between carbon atoms are modeled by the Tersoff empirical potential with parameters optimized for graphene and carbon nanotubes \cite{Lindsay2010, Lindsay2010a}.
This optimized Tersoff potential had been shown to appropriately reproduce phonon dispersions of graphene, and has been employed in several studies involving thermal transport \cite{Pereira2013, Pereira2013a, Xu2014, Mortazavi2014b, Mortazavi2015, Mortazavi2015a, Fan2015, Mortazavi2016} and mechanical properties \cite{Mortazavi2014c, Mortazavi2016, Mortazavi2016a} of graphene and graphene-like materials.

\begin{figure}[htb]
\centering
\includegraphics[width=\linewidth]{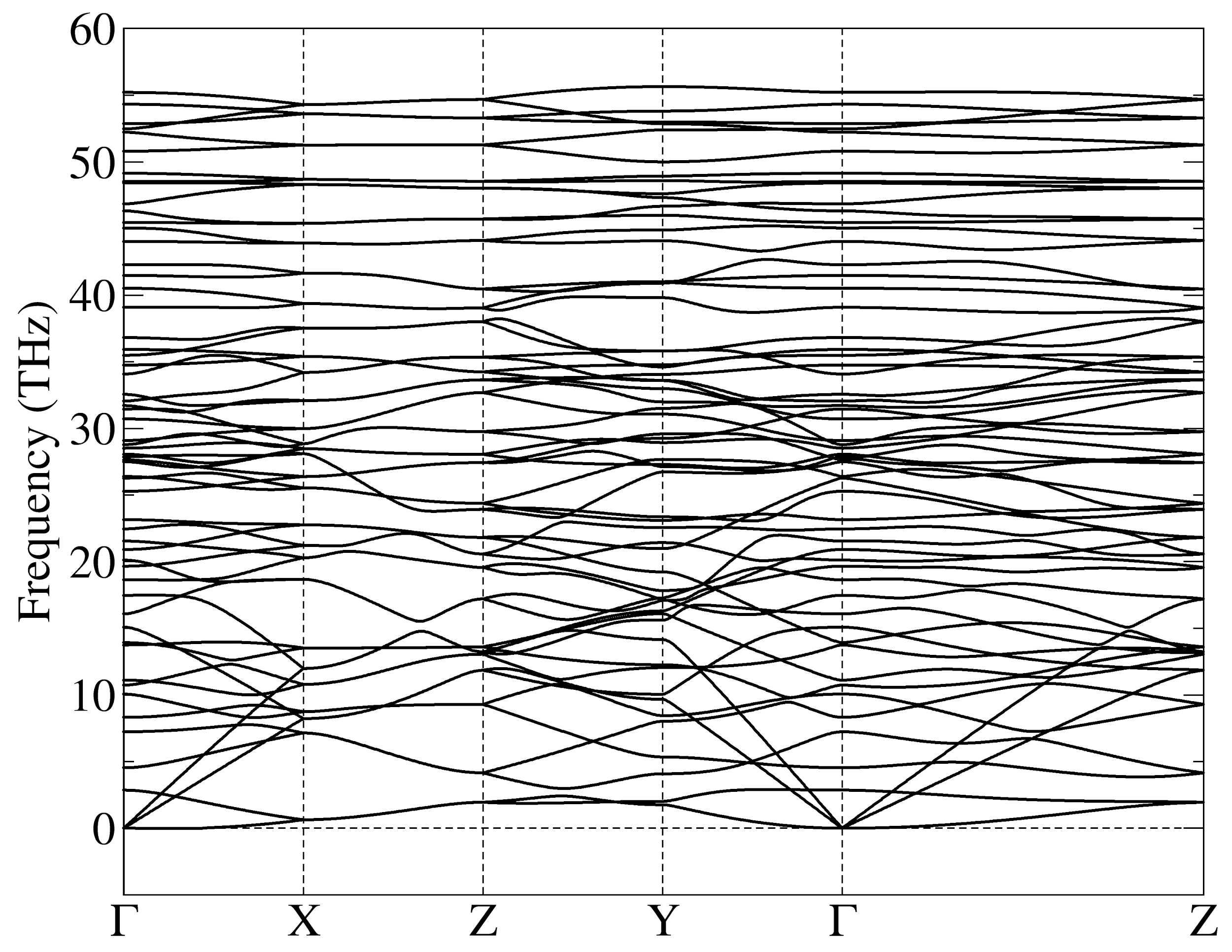}
\caption{Phonon dispersion for phagraphene modeled by the optimized Tersoff potential. The absence  of negative (imaginary) frequencies states the stability of the structure with the chosen potential parameters. The high symmetry points in reciprocal space coordinates are $\Gamma=(0,0,0)$,  X$=(0.5,0,0)$, Z$=(0.5,0.5,0)$, and Y$=(0,0.5,0)$.}
\label{fig02}
\end{figure}

In order to verify the accuracy of the optimized Tersoff potential in describing the atomic bonding structure of phagraphene we calculated its phonon dispersions via the lattice dynamics software GULP \cite{Gale1997, Gale2003}.
Fig. \ref{fig02} presents the phonon dispersions along high symmetry points of the Brillouin zone, obtained from the 20-atom unit cell shown in Fig. \ref{fig01}.
The absence of phonon modes with negative (imaginary) frequencies in the dispersion indicates that the crystal structure of phagraphene is stable when modeled by the chosen potential. 
In particular, of the three acoustic phonon modes, both in-plane modes present linear dispersions while the flexural mode presents a parabolic dispersion around the $\Gamma$ point.
Furthermore, the presence of high frequency modes ($\sim 50$ THz) such as in graphene is related to the high strength of the sp$^2$ chemical bonds between carbon atoms in phagraphene.

Next we performed room temperature molecular dynamics simulations with LAMMPS \cite{Plimpton1995}.
All simulations employed periodic boundary conditions along both in-plane directions and free boundary conditions in the perpendicular directions, such that our simulations would correspond to suspended phagraphene samples.
It is important to notice that the presence of 5- and 7-membered rings in the structure of phagraphene is similar to some topological defects found in graphene and may result in technical difficulties such as instabilities during simulations at finite temperature \cite{Mortazavi2014b, Mortazavi2015a}. 
To avoid this problem, in all molecular dynamics simulations the equations of motion were integrated with a relatively small time increment of $0.25$ fs.

NEMD simulations were performed to investigate the thermal conductivity of phagraphene along both in-plane directions, labeled armchair and zigzag in Fig. \ref{fig01}, independently. 
For each simulation, the system was divided in $22$ slabs along the sample length (the direction of heat transport), and atoms at both ends where fixed.
The whole structure, excluding fixed atoms, was then thermalized for $100$ ps at room temperature ($300$ K) with a Nos\'e-Hoover thermostat (NVT) \cite{Nose1984, Hoover1985}. 
In our simulations, a temperature gradient is imposed in the system via independent thermostats while the heat flux is calculated from the energy exchanged with the thermostats.
Therefore, the original thermostat was turned off, and two independent thermostats were coupled to the ``hot'' slab (first slab, $310$ K) and ``cold'' slab (22nd slab, $290$ K), giving rise to a temperature gradient along the system length.
All the atoms between the cold and hot regions evolve freely from thermostats, and the temperature of each slab can be calculated from the average kinetic energy of its atoms via the equipartition theorem.
The temperature of slab $n$ is given by
\begin{equation}
T_n = \frac{1}{3 N_n k_B} \sum m_i v_i^2,
\end{equation}
where $N_n$ is the number of atoms in the slab, $k_B$ is Boltzmann's constant, $m_i$ is the mass of atom $i$ and $v_i$ represents its velocity.
Meanwhile, the heat flux is given by the energy added to the hot slab and removed from the cold slab, such that it can be written as 
\begin{equation}
J_x=\frac{1}{A} \frac{dq}{dt}, 
\end{equation}
where  $A$ is the cross sectional area of the phagraphene sample and $\frac{dq}{dt}$ is the energy exchanged with the thermostats. 
The cross sectional area is given by the product of the sample width with the thickness of the phagraphene sheet.
All simulated samples had a width of approximately $8$ nm, while the thickness of phagraphene was taken to be $0.335$ nm, in analogy to the nominal value assumed for graphene.
After approximately $500$ ps of NEMD, the system achieves a steady-state where the average heat flux and the temperature gradient are stationary.
We then simulate the system for an additional time of at least $2.0$ ns and at most $4.0$ ns. 
Once the average heat flux and temperature gradient are stable, the conductivity ($\kappa$) can be calculated directly from Fourier law as
\begin{equation}
\kappa = \frac{\langle J_x \rangle }{\langle dT/dx \rangle },
\end{equation}
where $\langle \cdots \rangle$ indicates time averages.

For the evaluation of phagraphene's mechanical properties we also applied periodic boundary conditions along both in-plane directions, and the structures were equilibrated at $300$ K and zero-pressure via a Nos\'e-Hoover thermostat and barostat.
The loading of the structure was performed via deformation of the simulation box, which was increased along the loading direction by a constant engineering strain rate of $10^8$ s$^{-1}$ at every simulation time step.
In order to guarantee uniaxial stress conditions, a barostat was applied to the direction perpendicular to loading and set for zero-stress.
Virial stresses were calculated at each strain level to obtain the stress-strain curves presented in the next section, and the elastic modulus was obtained directly from the slope of the stress-strain curves.

\section{Results and discussion}

\subsection{Thermal conductivity}

\begin{figure}[htb]
\centering
\includegraphics[width=\linewidth]{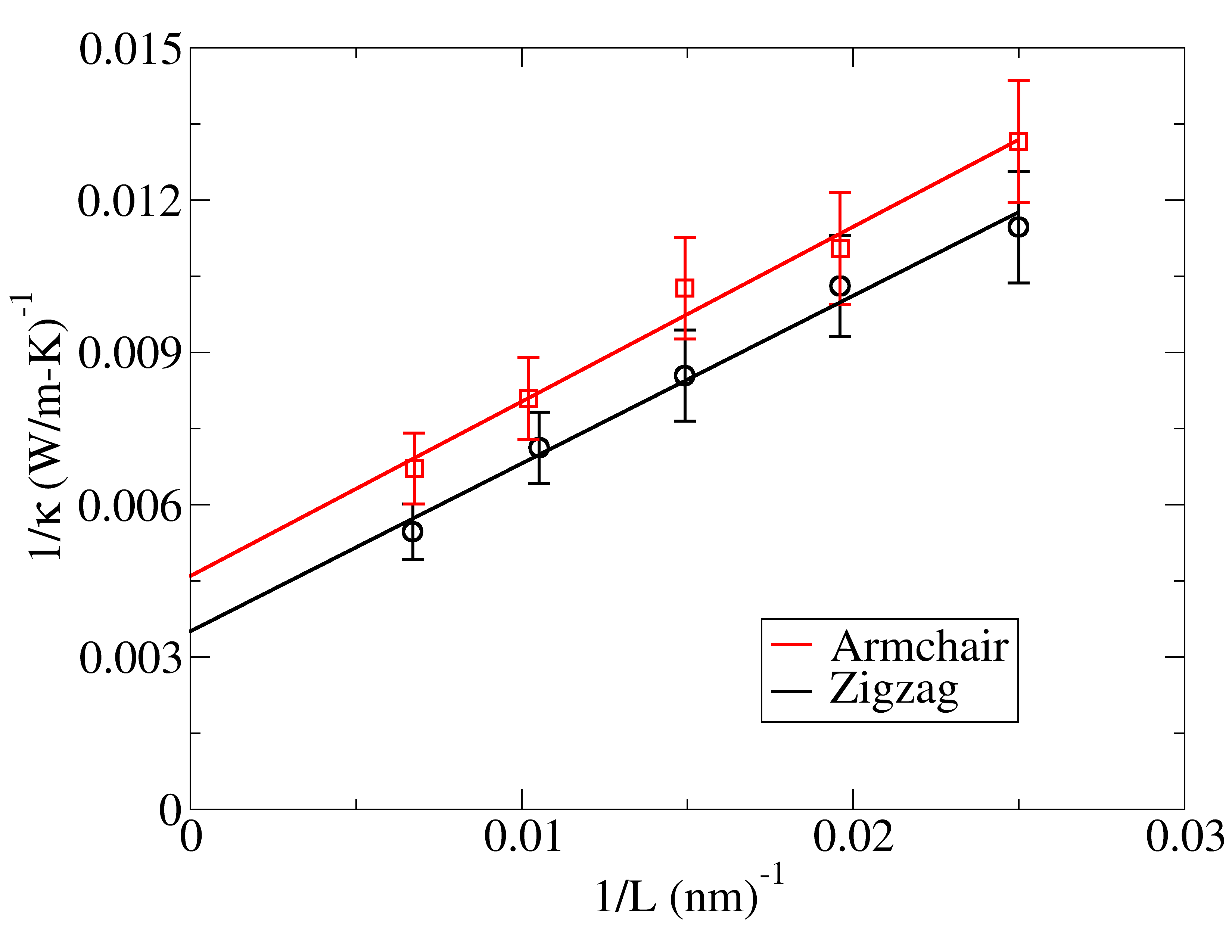}
\caption{Inverse thermal conductivity versus inverse length. The lines are best fits to the data from which we extract anisotropic thermal conductivities and effective phonon MFP for phagraphene.}
\label{fig03}
\end{figure}

The thermal conductivity is expected to present strong size-effects when the system length is smaller or comparable to the mean free path (MFP) of the heat carriers of the material \cite{Schelling2002,Sellan2010}. In the case of graphene, this behavior has been observed in NEMD simulations and even compared to experimental data \cite{Xu2014}.
In order to investigate this behavior in phagraphene, we performed NEMD simulations with cells of increasing length, and calculated their thermal conductivity along armchair and zigzag directions independently.
The size-dependence of $\kappa$ is shown in Fig. \ref{fig03}, which also predicts that the thermal conductivity along armchair and zigzag directions are not equal.
The in-plane thermal conductivity of phragraphene is anisotropic, with a higher conductivity along the zigzag direction indicated in Fig. \ref{fig01}. 
{This anisotropic behavior of $\kappa$ is in contrast to what has been observed for similar carbon-based materials such pristine graphene \cite{Pereira2013}, penta-graphene \cite{Xu2015a, Wang2016}, amorphized graphene \cite{Mortazavi2016} and nitrogenated holey graphene \cite{Mortazavi2016a}.}

In general, the size-dependent thermal conductivity is related to the intrinsic thermal conductivity of the material, via \cite{Schelling2002,Zhang2014}
\begin{equation}
\frac{1}{\kappa_L} = \frac{1}{\kappa} \left( 1 + \frac{\Lambda_{eff}}{L} \right),
\label{eq:nemd}
\end{equation}
where $\Lambda$ is an effective phonon MFP and $L$ is the system length. 
Within this description, the effective MFP corresponds to the length at which $\kappa_L$ equals $50$\% of $\kappa$.
Fitting the above expression to the NEMD data we can determine an intrinsic thermal conductivity of $218 \pm 20$ W/m-K and an effective phonon MFP of $74.9 \pm 5.6$ nm along the armchair direction, and $285 \pm 29$ W/m-K and $94.3 \pm 6.1$ nm along the zigzag direction.
In either direction, the thermal conductivity of phagraphene is at least one order of magnitude smaller when compared to $3050 \pm 100$ W/m-K obtained for pristine graphene with the same empirical potential \cite{Mortazavi2015}.
By fitting Eq. \ref{eq:nemd} to the graphene data set obtained in the previous study \cite{Mortazavi2015}, we can estimate its effective MFP to be $287.3 \pm 5.2$ nm.
The difference in $\kappa$ and $\Lambda_{eff}$ between graphene and phagraphene can be qualitatively understood in terms of the higher phonon scattering due to the presence of pentagons and heptagons in phagraphene. 
These 5- and 7-membered rings behave as defects in the hexagonal structure of pristine graphene, giving rise to scattering of heat carriers by defects in the case of phagraphene, therefore decreasing its thermal conductivity.
A similar decrease in thermal conductivity relative to graphene has also been predicted for penta-graphene via molecular dynamics simulations \cite{Xu2015a} and lattice dynamics \cite{Wang2016}.
The thermal conductivity of ultra-fine grained graphene was also predicted to be one order of magnitude smaller than pristine single-crystal graphene because of increased phonon scattering along grain boundaries \cite{Mortazavi2014b}. 

\begin{figure}[htb]
\centering
\includegraphics[width=\linewidth]{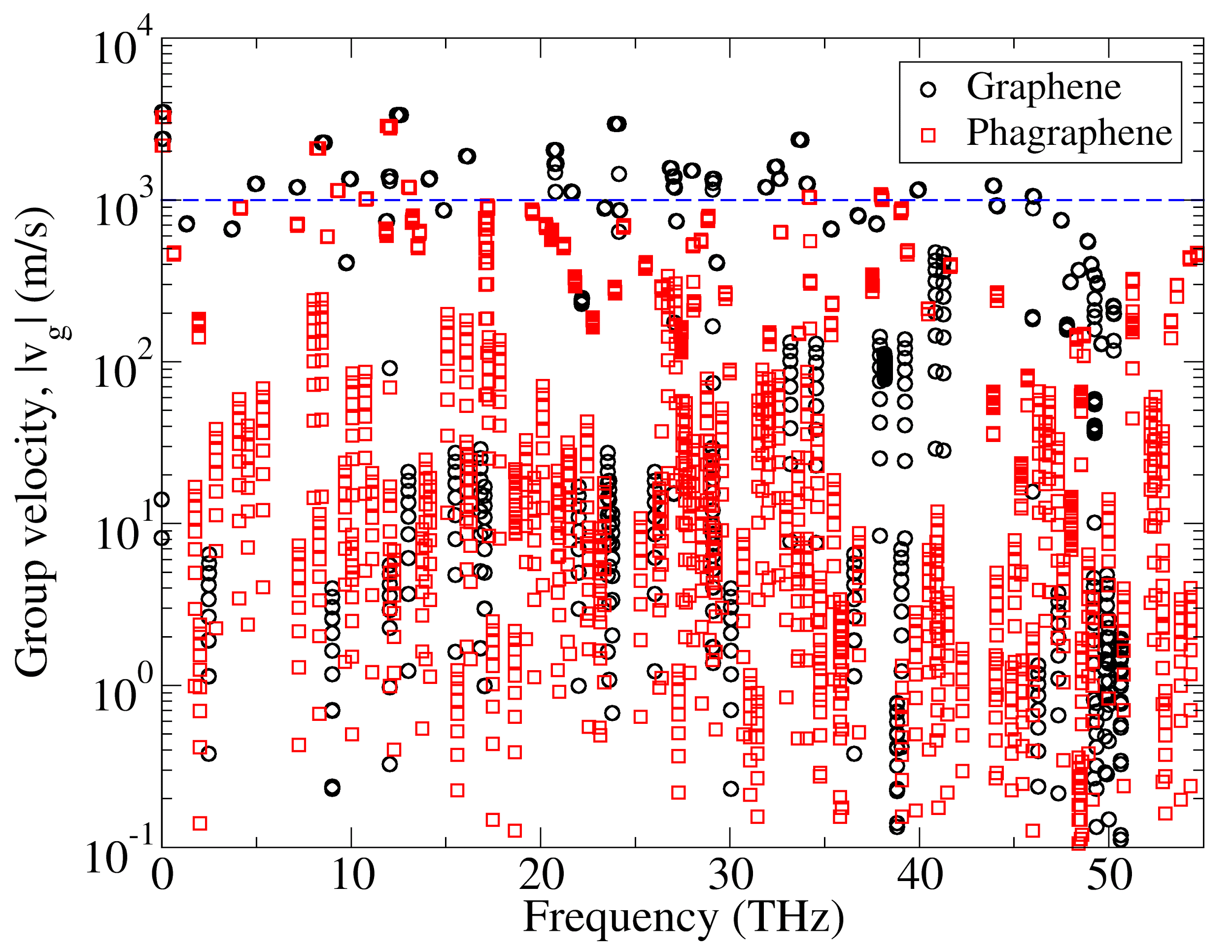}
\caption{Phonon group velocities as a function of frequency for graphene and phagraphene. The average group velocity for graphene is $780$ m/s while for phagraphene its only $125$ m/s. This difference alone explains part of the discrepancy between the thermal conductivities of the materials.}
\label{fig04}
\end{figure}

In order to better understand the difference between thermal conductivities of graphene and phagraphene, we have  also calculated phonon group velocities from the phonon dispersions of each material. The group velocities are shown in Fig. \ref{fig04} as a function of phonon frequencies, obtained from unit cells with approximately the same dimensions ($24$ atoms for graphene and $20$ atoms for phagraphene). Although the difference in group velocities for the acoustic modes is not so pronounced (notice the modes at $0$ THz), for higher frequency modes the group velocities are consistently larger for graphene, as evidenced by the dashed horizontal line.  Considering all phonon frequencies in Fig. \ref{fig04} we can calculate an average group velocity of $780$ m/s for graphene, while the equivalent average for phagraphene yields only $125$ m/s. 
This reduction in phonon group velocities is not the only reason for the lower thermal conductivity of phagraphene relative to graphene, but it provides further insight in to the origin of the conductivity reduction.
Interestingly, the lower thermal conductivity of phagraphene along with its predicted electronic properties suggests that phagraphene could be a better candidate than graphene in future carbon-based thermoelectric devices \cite{Wang2015a, Chang2012, Sevincli2013}.

\subsection{Mechanical Properties}

\begin{figure}[htb]
\centering
\includegraphics[width=\linewidth]{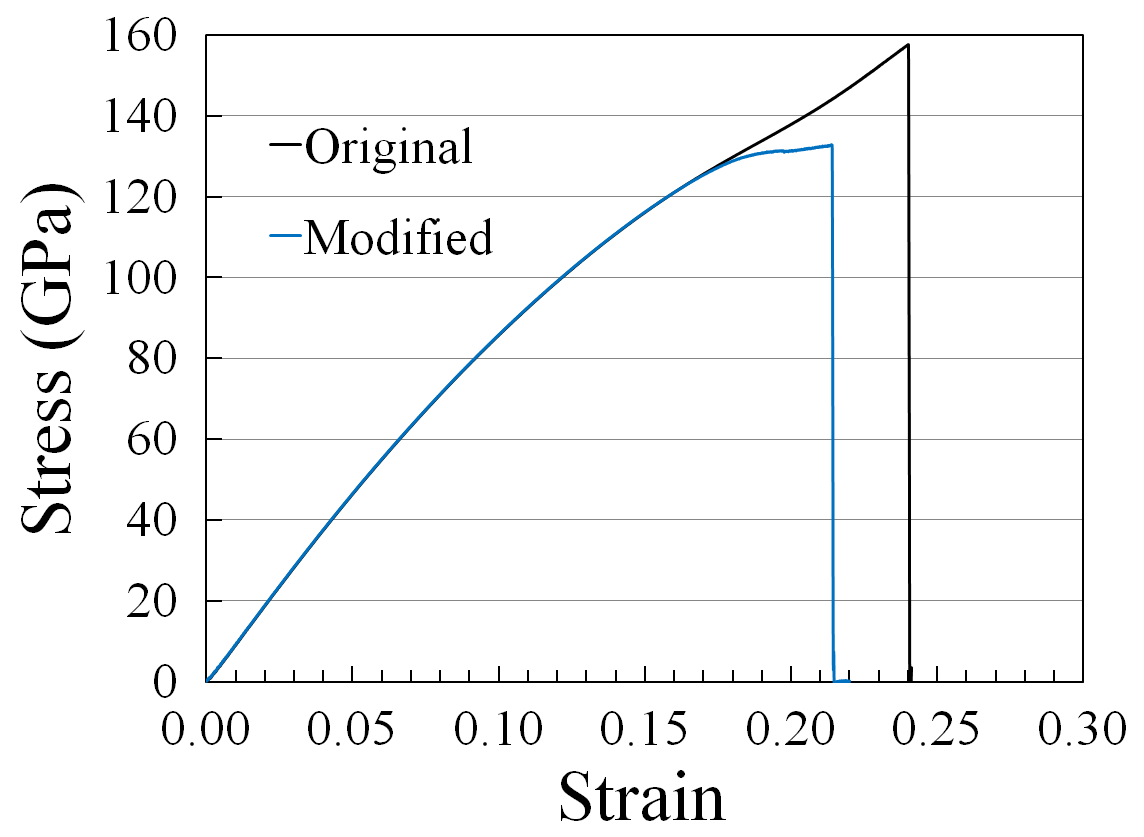}
\caption{Uniaxial stress-strain response of pristine graphene obtained with the optimized Tersoff potential and its cutoff-modified version. In the modified version the cutoff of the Tersoff potential was increased from $0.18$ nm to $0.2$ nm. }
\label{fig05}
\end{figure}

In Fig. \ref{fig05} we plot the stress-strain response of pristine graphene at room temperature using the parameter set of the optimized Tersoff potential \cite{Lindsay2010, Lindsay2010a}.
For strain levels higher than $0.2$ an unphysical strain hardening can be observed in stress values.
Similar behavior has previously been reported with the AIREBO \cite{Stuart2000} empirical potential in simulations investigating the tensile deformation of graphene \cite{Pei2010, Chen2015}.
In the case of AIREBO, the unphysical stress hardening can be removed by increasing the potential cutoff from $0.17$ nm to $0.2$ nm \cite{Chen2015}. 
In analogy to this situation, we also increased the cutoff of the optimized Tersoff potential from $0.18$ nm to $0.20$,  and the results obtained with this modification are also presented in Fig. \ref{fig05}. 
In accordance with the results using AIREBO, the stress-strain curves calculated with the optimized Tersoff potential  and its cutoff-modified version agree for strain levels up to $0.17$. 
However, for strain values larger than $0.17$, the unphysical behavior is no longer produced by the modified potential.
For pristine graphene, the modified potential yields an elastic modulus of $960 \pm 20$ GPa and a tensile strength of $132 \pm 3$ GPa.
These values are in remarkable agreement with experimental results of $1000 \pm 100$ GPa for elastic modulus and $130 \pm 10$ GPa for tensile strength of pristine graphene \cite{Lee2008}.

\begin{figure}[htb]
\centering
\includegraphics[width=\linewidth]{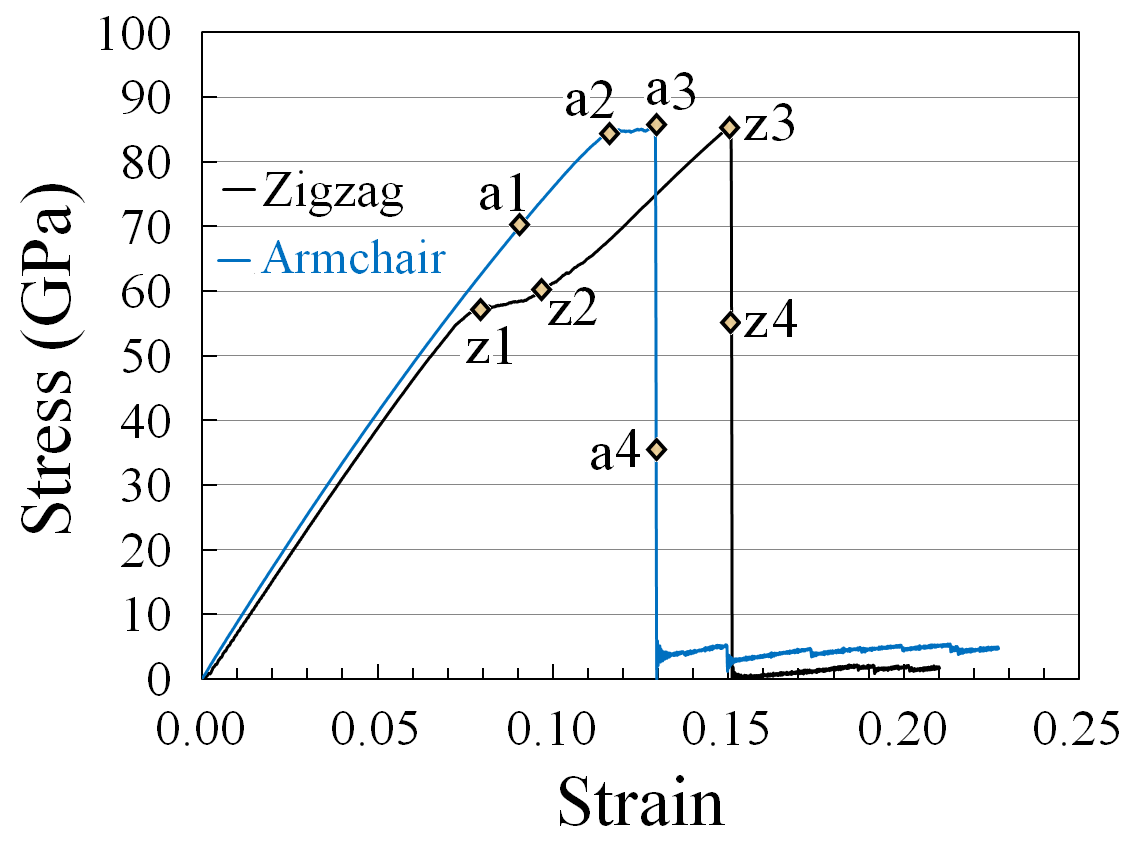}
\caption{Uniaxial stress-strain response of phagraphene along armchair and zigzag loading directions.}
\label{fig06}
\end{figure}

The stress-strain response curves of phagraphene along armchair and zigzag directions are presented in Fig. \ref{fig06}, where an anisotropy in the stress-strain curves along zigzag and armchair directions is apparent. 
Our simulations predict an elastic modulus of $870 \pm 15$ GPa along the armchair direction and $800 \pm 14$ GPa along the zigzag direction. 
We also predict a remarkable tensile rigidity of $85 \pm 2$ GPa for phagraphene, which is $64$\% of the value for pristine graphene { and larger than the corresponding values for graphynes \cite{Zhang2012}.}
Interestingly, our simulations show that the tensile strength of phagraphene is direction independent, as indicated by points a3 and z3 in Fig. \ref{fig06}.

For pristine graphene, the cutoff-modified optimized Tersoff potential predicts that the sample extends uniformly during uniaxial tensile loading and that it remains free of defects up to its tensile strength. 
At that point, the first debonding occurs between two adjacent carbon atoms. 
This results in the formation of a crack that rapidly grows and eventually leads to sample rupture. 
Therefore, for pristine graphene the initial void formation and subsequent failure occur at almost the same strain level, which suggests a brittle failure mechanism. 

\begin{figure*}[htb]
\centering
\includegraphics[width=0.7\linewidth]{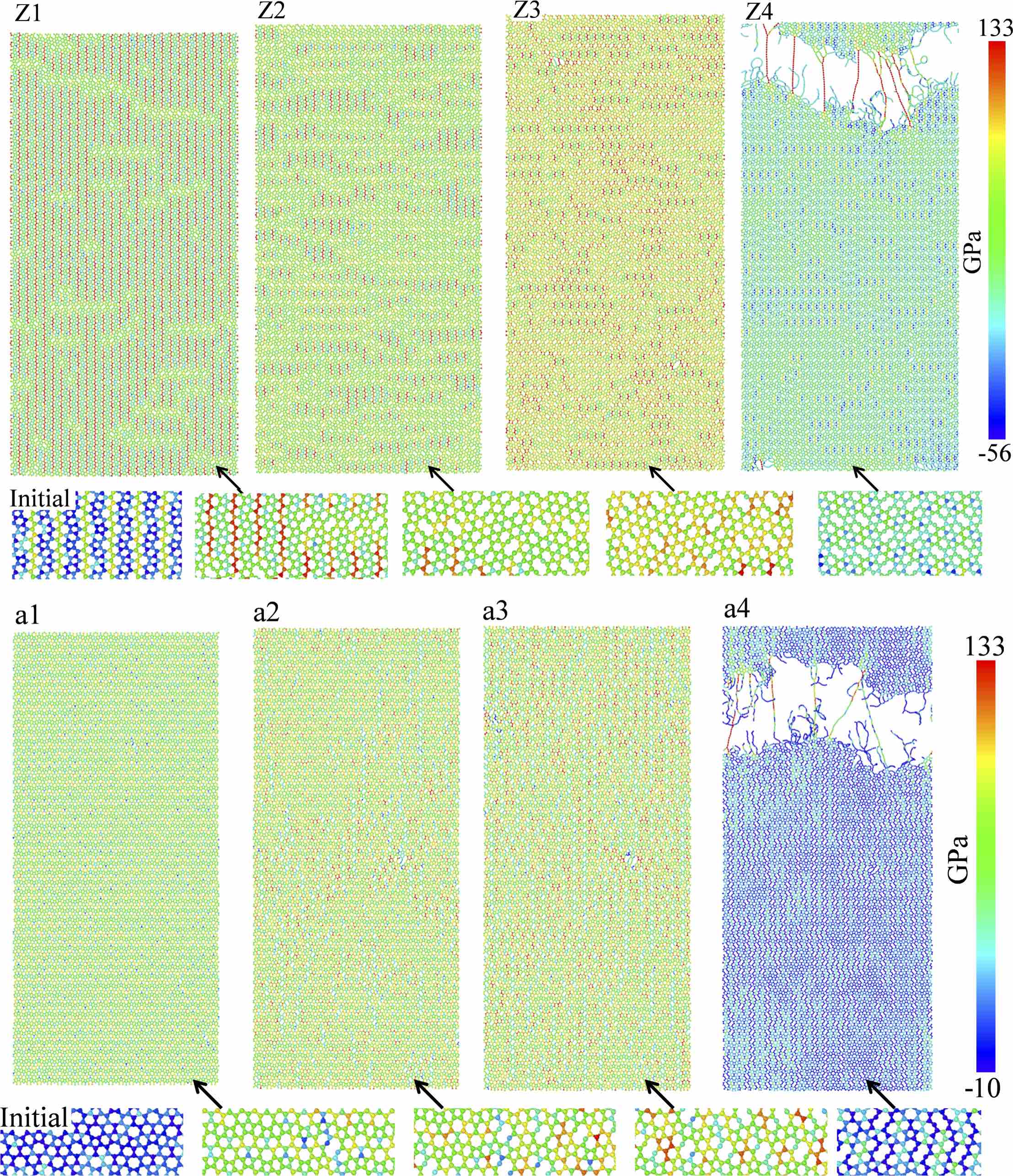}
\caption{Uniaxial deformation process of single-layer phagraphene at various engineering strain levels along zigzag (a) and armchair (z) loading directions, respectively. The strain and stress levels (z1-z4 and a1-a4) are depicted in Fig. \ref{fig06}. Images generated with OVITO\cite{Stukowski2010}.}
\label{fig07}
\end{figure*}

{In Fig. \ref{fig07}, the deformation and breakage process of phagraphene sheets along zigzag and armchair directions is depicted. 
For phagraphene under tensile loading along the zigzag direction, we observe the initial bond ruptures at strain levels of $0.07$ (Fig. \ref{fig07}z1). 
These debonding events do not happen in pentagon-heptagon pairs but occur in bonds that connect two hexagons consequently forming ten-membered rings from two carbon hexagons (Fig. \ref{fig07}z1). 
We observe that at the initial stages of loading, when the sample is stretched along the zigzag direction there are contraction and tension stresses in the sample. In this case we found that  the stress around pentagon and bonds connecting two hexagons are contraction and tension stress, respectively. These diverse stress conditions result in a non-uniform stress distribution and cause stress concentration in specific parts of the samples. Consequently the high tensile stress around bonds connecting two hexagons leads to local bond ruptures (Fig. \ref{fig07}z1 and Fig. \ref{fig07}z2) forming a more uniform stress distribution. 
These void initiations can be identified in stress-strain curves at the point in which the slope of the stress-strain curve decreases (point z1 in Fig. \ref{fig06}). 
Increasing the strain level along the zigzag direction, leads to more bond breakage and ten-membered rings extend through the entire sample (Fig. \ref{fig07}z2). 
After this point, the defect concentration along the sample remain almost unchanged and formation of new ten-membered rings slows down.
In this case, the density of defects is approximately uniform throughout the structure and the sample starts to extend uniformly.  
This process is distinguishable from the stress-strain response through an increase in the slope of stress-strain curve (Fig. \ref{fig07}z2 and z2 point in Fig. \ref{fig06}). 
The ultimate tensile strength happens at the point when the defect coalescence happens (Fig. \ref{fig07}z3) forming a crack that grows and leads to the sample rupture (Fig. \ref{fig07}z4). 
Meanwhile, our simulations reveal that when phagraphene is under tensile loading along the armchair direction, it remains defect-free up to high stages of the loading (Fig. \ref{fig07}a1). 
In this case, bonds connecting two hexagons are almost perpendicular to the loading direction and by increasing the strain level their bond length does not change considerably so the stress distribution is much more uniform in this case in comparison to the zigzag direction. A more uniform stress distribution therefore postpone the first bond breakage in the structure. Nevertheless, at strain levels higher than $0.1$, ten-membered rings start to form along the sample, which gradually extend throughout the structure (Fig. \ref{fig07}a2). 
Similarly to our results along the  zigzag direction, the coalescence of ten-membered rings results in the formation of a crack and subsequent rupture of the sample (Fig. \ref{fig07}a4). }

Unlike pristine graphene, in phagraphene the initial debonding and subsequent rupture happen at distinctly different strain levels, which suggests a ductile failure mechanics for phagraphene. It is worth noticing that, based on our simulations, phagraphene presents a higher ductile failure mechanism along the zigzag direction when compared to its armchair direction.
In this case we found that during the tensile deformation and up to sample rupture, phagraphene stretched along the zigzag direction can absorb around $14$\% more energy than when stretched along the armchair direction. 
In addition, along the zigzag direction phagraphene's failure strain is found to be $0.15$ while this is around $0.13$ for the armchair direction.

Table \ref{tab01} summarizes the predicted thermal transport and mechanical properties of phagraphene, along with the values obtained for pristine graphene with the same empirical potential.
While graphene presents anisotropic (in-plane) thermal transport and mechanical properties, phagraphene presents a clear anisotropy for the same properties.
Interestingly, while the thermal conductivity and the effective phonon MFP are predicted to be larger along the zigzag direction, the elastic modulus is predicted to be higher along the armchair direction.
Nonetheless, while $\kappa$ along the zigzag direction is $25$\% larger than along the armchair direction, the elastic modulus along the armchair direction is less than $10$\% larger than along the zigzag direction.
Furthermore, while we are very confident in our predicted anisotropic thermal conductivity, it is possible that the anisotropy observed in tensile strength is an artefact of our simulations, particularly due to the chosen empirical potential which was not explicitly fitted to structures such as phagraphene.

\begin{table}[htb]
\centering
\small
  \caption{\ Thermal transport and mechanical properties of graphene and phagraphene}
  \label{tab01}
  \begin{tabular*}{0.48\textwidth}{@{\extracolsep{\fill}}llll}
    \hline
     & Graphene & Phagraphene &   \\
    \hline
    & & Armchair & Zigzag \\
    \hline
    $\kappa$ (W/m-K) & $3050 \pm 100$ & $218 \pm 20$ & $285 \pm 29$ \\
    $\Lambda_{eff}$ (nm) &  $287.3 \pm 5.2$ & $74.9 \pm 5.6$ & $94.3 \pm 6.1$  \\
    Elast. Mod. (GPa) & $960 \pm 20$ & $870 \pm 15$ & $800 \pm 14$ \\
    Tens. Strength (GPa) & $132 \pm 3$ & $85 \pm 2$ & $85 \pm 2$  \\
    \hline
  \end{tabular*}
\end{table}

\section{Summary}

Thermal conductivity and mechanical response of phagraphene at room temperature were investigated via classical molecular dynamics simulations, along armchair and zigzag directions. 
NEMD simulations were performed to predict the thermal conductivity of phagraphene, which was found to be anisotropic for in-plane directions. 
We predict a room temperature thermal conductivity of $218 \pm 20$ W/m-K along the armchair direction and $285 \pm 29$ W/m-K along the zigzag direction, with effective MFP of $74.9 \pm 5.6$ nm and $94.3 \pm 6.1$ nm, respectively. 
The predicted in-plane thermal conductivities of phagraphene are one order of magnitude smaller than that of the pristine graphene, while effective phonon MFP are smaller by a factor of $3$.
Analysis of phonon group velocities also shows a significant reduction in this quantity for phagraphene in comparison to graphene.
We have also investigated the mechanical properties of phagraphene via uniaxial tensile tests. 
It was found that the cutoff modified optimized Tersoff potential predicts accurate mechanical properties for pristine graphene, in agreement with experiments. 
The elastic modulus of phagraphene along the armchair and zigzag directions are calculated to be $870 \pm 15$ GPa and $800 \pm 14$ GPa, respectively. 
In addition, the tensile strength of phagraphene is predicted to be around $85 \pm 2$ GPa,  independent of loading direction. 
Furthermore, the failure process of phagraphene was shown to be more ductile in comparison to pristine graphene. 
Our modeling results show that while the mechanical properties of phagraphene are comparable to graphene, its thermal conductivity is significantly smaller.
{Finally, our simulations show that the thermal conductivity and the elastic modulus of phagraphene are direction dependent, presenting an in-plane anisotropy not observed in the case of other two-dimensional carbon-based materials.}
The lower thermal conductivity coupled with its predicted electronic properties makes phagraphene a more promising candidate for carbon-based thermoelectric devices than graphene. 

\section*{Acknowledgments}
The authors thank Raphael Matozo Tromer for a critical reading of the manuscript.
LFCP acknowledges financial support from Brazilian government agency CAPES for project ``Physical properties of nanostructured materials'' {(grant number 3195/2014)} via its Science Without Borders program.
BM and TR greatly acknowledge the financial support by European Research Council for COMBAT project {(grant number 615132).}

%%%END OF MAIN TEXT%%%

%The \balance command can be used to balance the columns on the final page if desired. It should be placed anywhere within the first column of the last page.

%\balance

%If notes are included in your references you can change the title from 'References' to 'Notes and references' using the following command:
%\renewcommand\refname{Notes and references}

%%%REFERENCES%%%
\bibliography{/Users/pereira/Dropbox/Documents/library} %You need to replace "rsc" on this line with the name of your .bib file
\bibliographystyle{rsc} %the RSC's .bst file

\end{document}